\documentclass[aps,pra,twocolumn,groupedaddress,notitlepage=,
floatfix,superscriptaddress]{revtex4-1}

\pdfoutput=1
\usepackage{graphicx,graphics,epsfig,subfigure,times,bm,bbm,amssymb,amsmath,amsthm,mathrsfs,MnSymbol}
\usepackage{gensymb}
\usepackage{amsfonts}
\usepackage{float}
\usepackage[matrix,frame,arrow]{xypic}
\usepackage[pdfstartview=FitH]{hyperref}
\hypersetup{
	colorlinks=true,       
	linkcolor=red,          
	citecolor=blue,        
	filecolor=magenta,      
	urlcolor=blue,           
	runcolor=cyan
}
\usepackage[pdftex]{color}
\usepackage{braket}  
\usepackage{enumerate}
\usepackage[normalem]{ulem}
\usepackage[usenames,dvipsnames]{xcolor}
\usepackage{multirow}
\usepackage{mathtools}
\usepackage{bbm}
\usepackage{titletoc}

\definecolor{orange}{rgb}{1,0.5,0}

\newcommand{\RNum}[1]{\uppercase\expandafter{\romannumeral #1\relax}}

\newcommand{\kb}[1]{\langle #1 \rangle}

\newcommand{\ignore}[1]{}
\usepackage{geometry}

\ignore{
	\documentclass[eprintnumbers,amsmath,amssymb,onecolumn,a4paper,caption ]{article}
	\usepackage{amsfonts}
	\usepackage{amssymb}
	\usepackage{mathrsfs}
	\usepackage{mathbbold}
	\usepackage{bbm}
	\usepackage{mathrsfs}
	\usepackage{dcolumn}
	\usepackage{bm}
	\usepackage{times,epsfig,amssymb,amsmath}
	\usepackage{float}
	\usepackage{subfigure}
	\usepackage{geometry}\geometry{left=2.5cm,right=2.5cm,top=3cm,bottom=3cm}

	\usepackage{color}

}

\begin{document}
	\title{Localization of Rung Pairs in Hard-core Bose-Hubbard Ladder}
	
	\author{Shang-Shu~Li}
	\altaffiliation[]{These authors contributed equally to this work.}
	\affiliation{Beijing National laboratory for Condensed Matter Physics, Institute of Physics, Chinese Academy of Sciences, Beijing 100190, China}
	\affiliation{School of Physical Sciences, University of Chinese Academy of Sciences, Beijing 100190, China}
	
	\author{Zi-Yong~Ge}
	\altaffiliation[]{These authors contributed equally to this work.}
	\affiliation{Beijing National laboratory for Condensed Matter Physics, Institute of Physics, Chinese Academy of Sciences, Beijing 100190, China}
	\affiliation{School of Physical Sciences, University of Chinese Academy of Sciences, Beijing 100190, China}
	
	\author{Heng~Fan}
	\email{hfan@iphy.ac.cn}
	\affiliation{Beijing National laboratory for Condensed Matter Physics, Institute of Physics, Chinese Academy of Sciences, Beijing 100190, China}
	\affiliation{School of Physical Sciences, University of Chinese Academy of Sciences, Beijing 100190, China}
	\affiliation{CAS Center for Excellence in Topological Quantum Computation, UCAS, Beijing 100190, China}
	\affiliation{Beijing Academy of Quantum Information Sciences, Beijing 100193, China}

	\begin{abstract}
Quantum simulation in experiments of many-body systems may bring new phenomena which are not well studied theoretically.    	
Motivated by a recent work of quantum simulation on a superconducting ladder circuit, we investigate the rung-pair localization of the Bose-Hubbard ladder model without quenched disorder. Our results show that, in the hard-core limit, there exists a rung-pair localization both at the edges and in the bulk. Using center-of-mass frame, the two-particle system can be mapped to an effective single-particle system with an approximate sub-lattice symmetry.
Under the condition of hard-core limit, the effective system is forced to have a defect at the left edge leading to a zero-energy flat band, which is the origin of the rung-pair localization.
We also study the multi-particle dynamics of the Bose-Hubbard ladder model, which is beyond the single-particle picture. In this case, we find that the localization can still survive despite of the existence of interaction between the pairs. Moreover, the numerical results show that the entanglement entropy exhibits a long-time logarithmic growth and the saturated values satisfy a volume law. This phenomenon implies that the interaction plays an important role during the dynamics, although it cannot break the localization. Our results reveal another interesting type of disorder-free localization related to a zero-energy flat band, which is induced by on-site interaction and specific lattice symmetry.

\end{abstract}
\maketitle{\tiny}

\section{Introduction}
Localization is a fundamental concept in condensed matter physics, which is closely related to the transports~\cite{beenakker}, non-equilibrium dynamics~\cite{polkov} and topology~\cite{hasan,qi}.
The localization can emerge in various systems.
In the free-fermion systems, the single-particle wave functions can be localized with the presence of sufficient impurity scatterings known as Anderson localization (AL)~\cite{anderson}.
Such disorder induced localization can be extended to interacting system dubbed many-body localization (MBL)~\cite{bardarson,pal,nandkishore,huse,zangara,chandran,vosk} as long as the disorder is strong enough.
In recent two decades, MBL has attracted many interests due to its novel properties, for instance, the violation of eigenstate thermalization hypothesis (ETH)~\cite{deutsch,srednicki,rigol} and long-time logarithmic growth of entanglement entropy.
Additionally, the localization can also exist in some disorder-free systems and have rich physics.
For example, there exists quasi-localization~\cite{schiulaz1,grover,hickey,schiulaz,yao,li,barbiero} in some translation invariant system which is induced by purely interacting effect.
In topological systems~\cite{kitaev,su}, localization can live at the boundaries protected by the bulk topology.
Other instances can be found in the locally constrained systems~\cite{Smith2017_1,Smith2017_2,Chen2018,brenes} due to the presence of superselection sectors and some flat-band systems~\cite{mukherjee,kuno,torma,takayoshi, mondaini,yang, kobayashi}, in which the localization is related to distractive interference of particle hopping.

In a recent quantum simulation experiment~\cite{ye}, a novel disorder-free localization phenomenon is observed,
in which some of our coauthors are involved.
Using a 20-qubit superconducting quantum simulator~\cite{ye,yan,song,xu}, we construct a Bose-Hubbard ladder~\cite{tschischik,keles} with equal inter- and intra-leg  hopping strength. From the dynamics of two particles, a special localization of the edge rung pair was observed, while the bulk rung pair exhibits a linear propagating.
It has been shown  that this localization is induced by large on-site interactions and specific lattice symmetries.
Nevertheless, there still exist two open questions:
i) Why and how the on-site interaction leading to this  localization.
ii) Whether the localization can still survive when there are many particles in the system.

In this paper, we theoretically investigate this localization of rung pairs in Bose-Hubbard ladder model. Our results reveal that,  the  localization of the single rung pair can emerge not only at the edges but also in the bulk.  In the center-of-mass frame,  two-particle system can be mapped into an effective single-particle Hamiltonian, and the corresponding spectrum is obtained. We find that there exist a zero-energy flat band in hard-core limit, which is the origin of the localization.
 Furthermore, we also study the dynamics of the multiple rung pairs, where the system cannot be described by single-particle Hamiltonian. It is shown that the localization can  still exist even in the case of half filling. In addition, the spreading  of entanglement entropy can display a long-time logarithmic growth, which is beyond single-particle picture.
 Thus, in the case of multiple rung pairs, the corresponding localization is not a single-particle phenomenon, since the interaction plays an important role during the quench dynamics.

The paper is organized as follows.  In Sec.~\ref{sec2}, we give a brief introduction to the model Hamiltonian. In Sec.~\ref{sec3}, we display the numerical results for single rung pair dynamics with various model parameter.  The mechanism of this localization in Sec.~\ref{sec4} is revealed by solving the spectrum of two-particle system. In Sec.~\ref{sec5}, we calculate numerically the dynamics of multiple rung pairs. Sec.~\ref{sec6} provides the summary and discussion. More theoretical and numerical details are given in Appendix.

\begin{figure}[t]
	\centering
	\includegraphics[width=8cm, height=3.5cm]{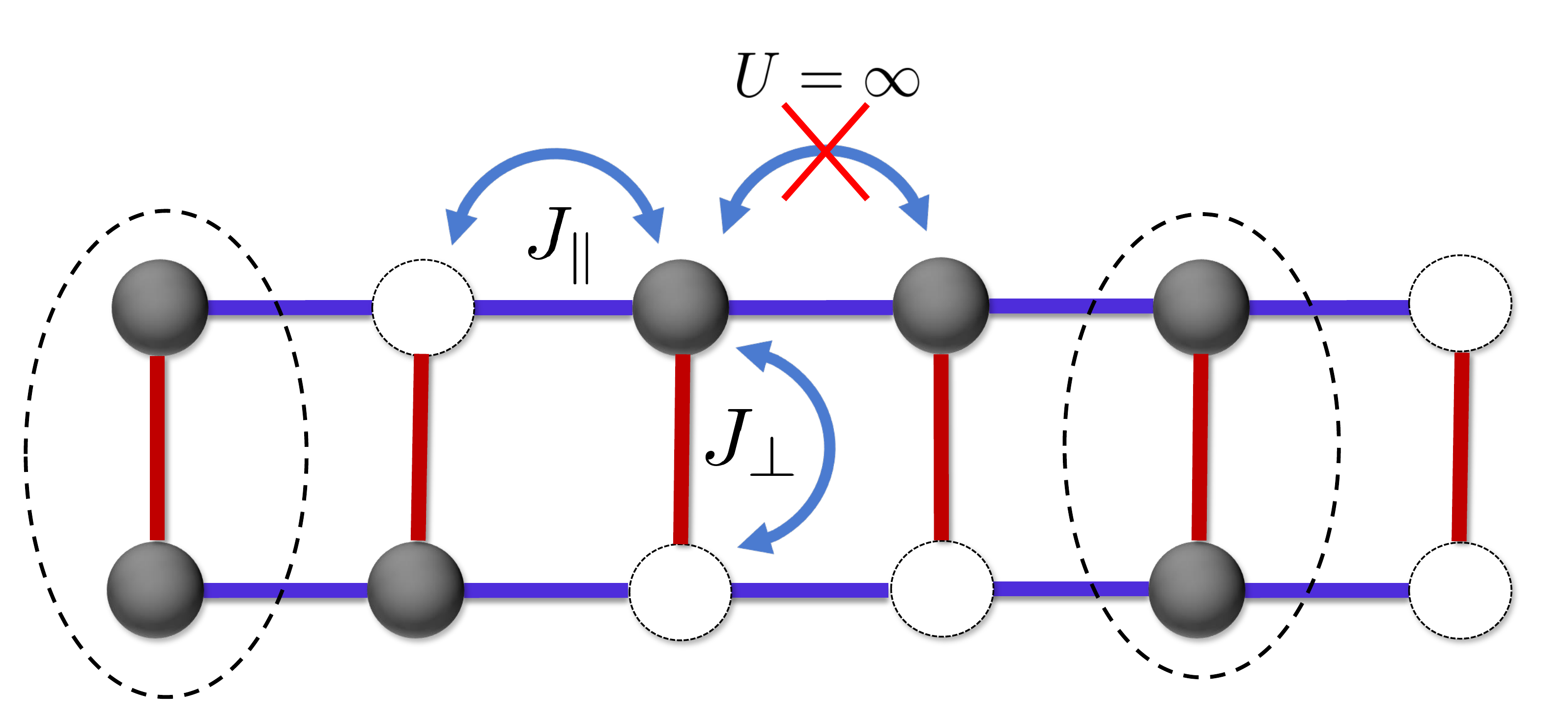}
	\caption{Sketch of Bose-Hubbard ladder.  The particles can hop through or between the legs with strength $J_{\parallel}$ and $J_{\perp}$, respectively. Double occupancy on a single site is forbidden in the hard-core limit. The dashed circles represent the rung pairs.
	}
	\label{fig:shiyi}
\end{figure}

\section{The Model}\label{sec2}
We consider the Bose-Hubbard ladder model with the Hamiltonian,
\begin{equation}\label{eq:H}
\begin{aligned}
\hat H &=J_{\parallel} \sum_{j, \nu}(\hat{a}_{j, \nu}^{\dagger} \hat{a}_{j+1, \nu}+ \text{H.c.})+J_{\perp} \sum_{j}(\hat{a}_{j,l_1}^{\dagger} \hat{a}_{j, l_2}+\text{H.c.})\\
&+\frac{U}{2} \sum_{j, \nu} \hat{n}_{j, \nu}(\hat{n}_{j, \nu}-1),
\end{aligned}
\end{equation}
where $\hat{a}^{\dagger}_{j, \nu}$ ($\hat{a}_{j, \nu}$) is bosonic creation (annihilation) operator, $\hat{n}_{j, \nu}=\hat{a}_{j,\nu}^{\dagger}\hat{a}_{j,\nu}$ is number operator of the boson, $j$ labels the rung indice, $\nu=l_1,l_2$ denotes two legs, the coefficient $J_{\parallel}$ and $J_{\perp}$ are intra- and inter-leg hopping strengthes respectively, and $U$ is the on-site interaction strength.  When $U\rightarrow \infty$, the boson is in the hard-core limit,  where a single site cannot be occupied by more than one boson, see Fig.~\ref{fig:shiyi}.
In this case, the Hamiltonian $\hat H$ is equivalent to a spin-$\frac{1}{2}$ ladder with XX coupling~\cite{ye}. For the system size, we use $L$ and $N\equiv L/2$ to denote the number of sites and rungs respectively.

Now we discuss the symmetries of the Hamiltonian (\ref{eq:H}).
Firstly, there is a global $U(1)$ symmetry, so that the particle number are conserved

\begin{equation}
 [\sum_{j, \nu}{\hat{n}_{j, \nu}}, \hat H]=0.
 \end{equation}
 Another  one is the space-reflection symmetry between the legs $l_1$ and $l_2$, i.e.,
 \begin{equation}\label{Ssym}
 [\hat S, \hat H]=0,
 \end{equation}
where  the symmetry transformation operator $\hat S$  satisfies $\hat S \hat a_{j, l_1}^{\dagger} \hat S^\dagger = \hat a_{j, l_2}^{\dagger}$ and $\hat S \hat a_{j, l_2}^{\dagger} \hat S^\dagger = \hat a_{j, l_1}^{\dagger}$.
We can define two projecting operators
\begin{equation}
\hat{P}^{\pm}=\frac{1}{2}(\hat{I}\pm\hat{S}),
\end{equation}
 which divide the Hamiltonian $\hat H$ into two subspaces $\hat H^\pm=\hat P^\pm \hat H \hat P^\pm$ with $\pm$ parities. These two symmetries are useful for constructing two-particle Hilbert space and deriving the spectrum in Sec.~\ref{sec4}.

\section{Single rung pair localization}\label{sec3}
In this section, we study the dynamics of Bose-Hubbard ladder with single rung occupied  under the open boundary condition.
The initial states $\ket{\psi_0}$ are chosen as $\ket{\psi_0} =\ket{\text{SRP}_{j_0}}\equiv\hat a_{j_0, l_1}^{\dagger}\hat a_{j_0, l_2}^{\dagger}\ket{\text{Vac}}$, where $\ket{\text{Vac}}$ is vacuum state of the boson satisfying $\hat a_{j, \nu}\ket{\text{Vac}}=0$. That is, the initial state only contains one  rung pair at the rung $j_0$.  Then we consider the quench dynamics under the  Hamiltonian 	$\hat H$.
 Another notation $\hat{n}_{i}$ for  $\hat{n}_{j,\nu}$ in one-dimensional chain representation of the ladder can be found in Appendix ~\ref{appdB}.

\begin{figure}[t]
	\centering
	\includegraphics[width=0.48\textwidth=]{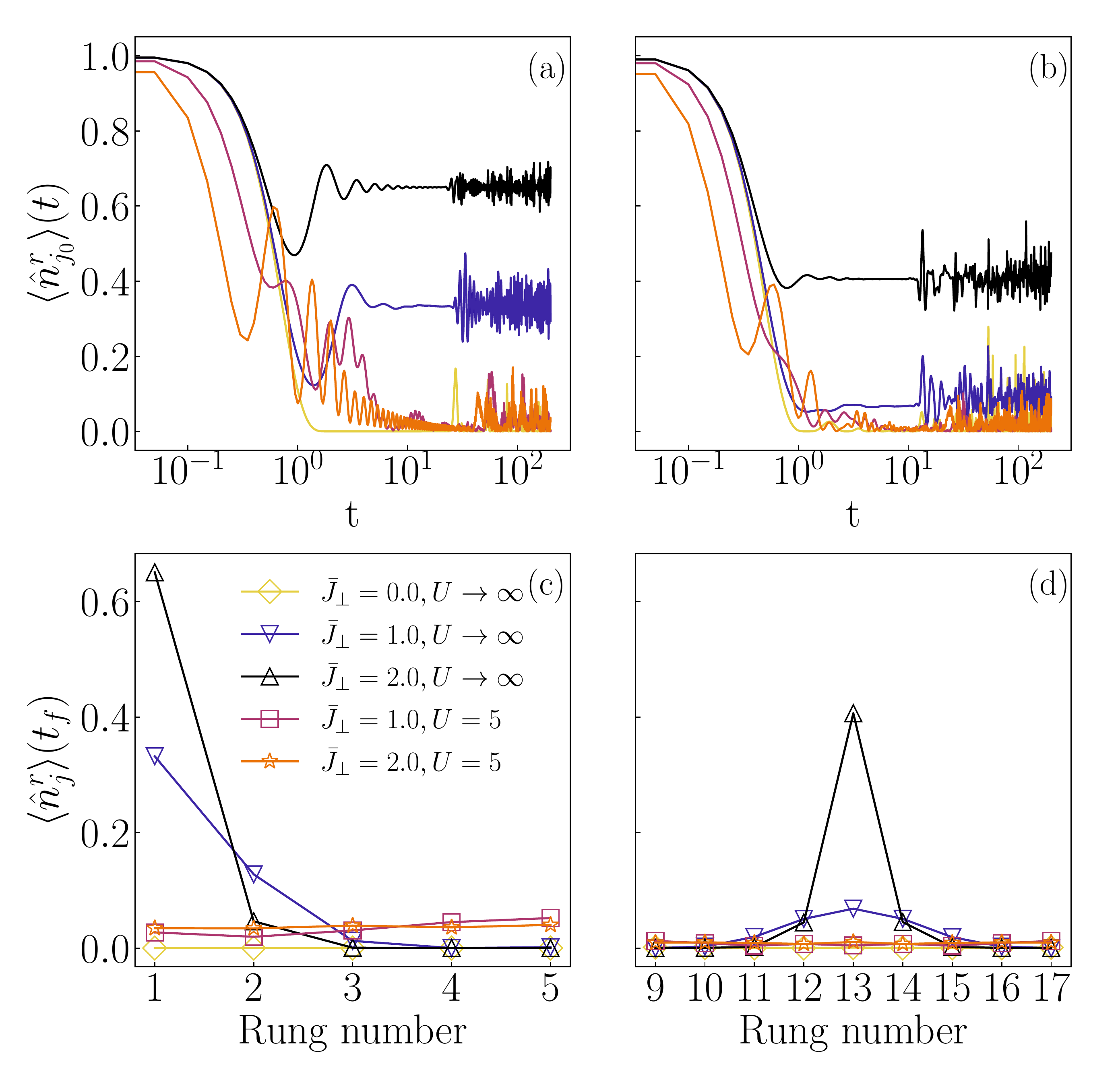}
	\caption{The dynamics of the single rung pair with $J_{\parallel}=1.0$ and system size $L=50$. The time evolution of  $\kb{\hat{n}^r_{j_0}(t)}$ with initial rung pairs (a) at the edge, i.e., $j_0=1$ and (b) the central, i.e., $j_0=13$.
		The corresponding occupancy probabilities can stabilize at non-zero values after a long time with $U\rightarrow\infty$ where the oscillations originate from the reflection of the particles when approaching the boundaries.
		The probability distributions of rung pairs $\kb{\hat{n}^r_{j}(t_f)}$ around the initial rungs pairs (c) at the edge and (d) the central, respectively.
		Here, $t_f$ is the time where the system has been approached at a steady state.}
	
	\label{fig:fig2}
\end{figure}

Firstly, we calculate the time evolution of the occupancy probability of the rung pair $\kb{\hat{n}^r_{j}(t)}:=\bra{\psi(t)}\hat{n}_{j,l_1}\hat{n}_{j, l_2} \ket{\psi(t)}$, where $\ket{\psi(t)}=e^{-i\hat H t}\ket{\psi_{0}}$. Here, $\kb{\hat{n}^r_{j_0}(t)}$, which represents the occupancy probability of the initial rung pair,  can also be considered as the Loschmidt echo~\cite{peres}.
In Figs.~\ref{fig:fig2}(a,b), we show the time evolution of $\kb{\hat{n}^r_{j_0}(t)}$ for different reduced inter-leg hopping strengh $\bar{J}_{\perp}\equiv J_{\perp}/J_{\parallel}$ with rung pairs initially at the boundary and central, respectively. It is shown that rung pairs can localize both at the edges and in the bulk in the hard-core limit as long as $J_{\perp}\neq0$.
However, in the case of small $U$, the  rung pairs can hardly localize. Furthermore, the localization strength of rung pairs is $\bar J_\perp$-dependent,
where it becomes stronger with the increase of reduced inter-leg hopping strength.
Then, we present the probability distributions of rung pairs $\kb{\hat{n}^r_{j}(t)}$ in the vicinity of initial rung, when the system approach a stead state, see Figs.~\ref{fig:fig2}(c,d).
Another property is that
the localization displays an edge effect, i.e., the boundary localization is stronger than the bulk one.
We note that the work in Ref.~\cite{ye} is a particular case with $\bar{J}_{\perp}=1$, where the bulk rung-pair localization is too weak to be observed experimentally.

\section{Spectrum of two particles}\label{sec4}
To uncover the mechanism of the rung-pair localization in Bose-Hubbard ladder model, we solve the Hamiltonian~(\ref{eq:H}) in two-particle subspace ~\cite{nguenang,zhang} and obtain the spectrum.
For two-particle problem, it is convenient to choose center-of-mass frame and  periodic boundary condition, i.e. $\hat{a}_{j, \nu}=\hat{a}_{j+N, \nu}$. Thus, the system is translation invariant with center-of-mass momentum being a good quantum number.
In addition, due to the parity symmetry (see Eq.~(\ref{Ssym})), we can choose the two-particle basis with certain  parity. For simplicity, we choose the system with odd rung number. Then the translation invariant bases with $+$ parity can be written as
\begin{equation}\label{basis}
\begin{aligned}
&|\varphi_{r,A}(K)\rangle \!=\!\frac{1}{\sqrt{2 N}} \sum_{j} e^{i K(j+r / 2)}(|1_{j} 1_{j+r}\rangle_{l_1}\!+\!|1_{j} 1_{j+r}\rangle_{l_2}), \\
&|\varphi_{r,B}(K)\rangle \!=\!\frac{1}{\sqrt{2N+2N\delta_{r,0}}} \sum_{j} e^{i K(j+r / 2)}(|1_{j}\rangle_{l_1}|1_{j+r}\rangle_{l_2}\\
&\ \ \ \ \ \ \ \ \ \ \ \ \ \ \ \ \ \ \ \ \ \  +\!|1_{j}\rangle_{l_2}|1_{j+r}\rangle_{l_1}),
\end{aligned}
\end{equation}
where $0\leq r\leq \frac{N-1}{2}$ is integer, $K=\frac{2\pi n}{N}$  is  the center-of-mass momentum with $n=0,1, \ldots, N-1$, and $|1_{j}\rangle_{\nu}\equiv\hat a_{j, \nu}^\dagger\ket{\text{Vac}}$.  Here, we have identified $\ket{1_j1_j}$ with the Fock state $\ket{2_j}$. Hence, the arbitrary two-particle eigenstate of $\hat H$ with momentum $K$ can be expand in this basis as
\begin{equation}\label{eq:psi}
|\psi(K)\rangle=\sum_{r } C^{K}_{A}(r)|\varphi_{r, A}(K)\rangle+\sum_{r} C^{K}_{B}(r)|\varphi_{r, B}(K)\rangle.
\end{equation}
From the Schr\"odinger's equation
\begin{equation}\label{eq:scheq}
\hat H\ket{\psi(K)}=\varepsilon_K\ket{\psi(K)}
\end{equation}
we can obtain an effective Hamiltonian of $\hat H$ in the basis of Eq.~(\ref{basis}) [See  Appendix.~\ref{appdA}], which reads
 \begin{equation}\label{Heff}
 \begin{aligned}
\hat H_{\mathrm{eff}} &=\sum_{r, \mu} Q_{r}^{K}|r\rangle_{\mu} {_{\mu}}\langle r+1|+2 J_{\perp} \sum_{r} | r\rangle_{A} {_B}\langle r|+\text{H.c.} \\
&+U|0\rangle_{A} {_A}\langle 0| + \sum_{\mu}(-1)^{n} Q_{N_{0}}^{K} | N_{0}\rangle_{\mu} {_{\mu}}\langle N_{0}| ,
\end{aligned}
 \end{equation}
where $ \mu=A, B , N_0\equiv \frac{N-1}{2}$ and $| r\rangle_{\mu} $ is alias of basis $|\varphi_{r,\mu}(K)\rangle$. The factors $Q^K_r$ are give by
\begin{equation}
 Q^{K}_{0}=2 \sqrt{2} J_{\|} \cos (K / 2), \quad Q^{K}_{r\geq 1}=2 J_{\|} \cos (K / 2).
 \end{equation}
This is a non-interactional Hamiltonian and has an approximate sub-lattice symmetry, i.e., the system is invariant under the exchange of $A,B$ sub-lattices except at left boundary.
The dispersion of $\hat H_{\mathrm{eff}}$ is calculated as [See Appendix.~\ref{appdA}]
\begin{equation}\label{sol}
\varepsilon_{K}^{\pm}(k)=4 J_{\|} \cos( K / 2) \cos (k) \pm 2 J_{\perp},
\end{equation}
where $k\in(0,\pi)$ is the relative momentum of two particles satisfying equation
\begin{equation}
\sin [k(N_0+1)]-(-1)^n \sin (k N_0)=0.
\end{equation}
There are two separated bands with the gap $4J_{\perp}$.
In addition, the winding  number is zero for arbitrary $J_{\perp}$ and $K$ indicating that the system is always topologically trivial.

 	\begin{figure}[t]
 	\centering
 	\includegraphics[width=0.48\textwidth]{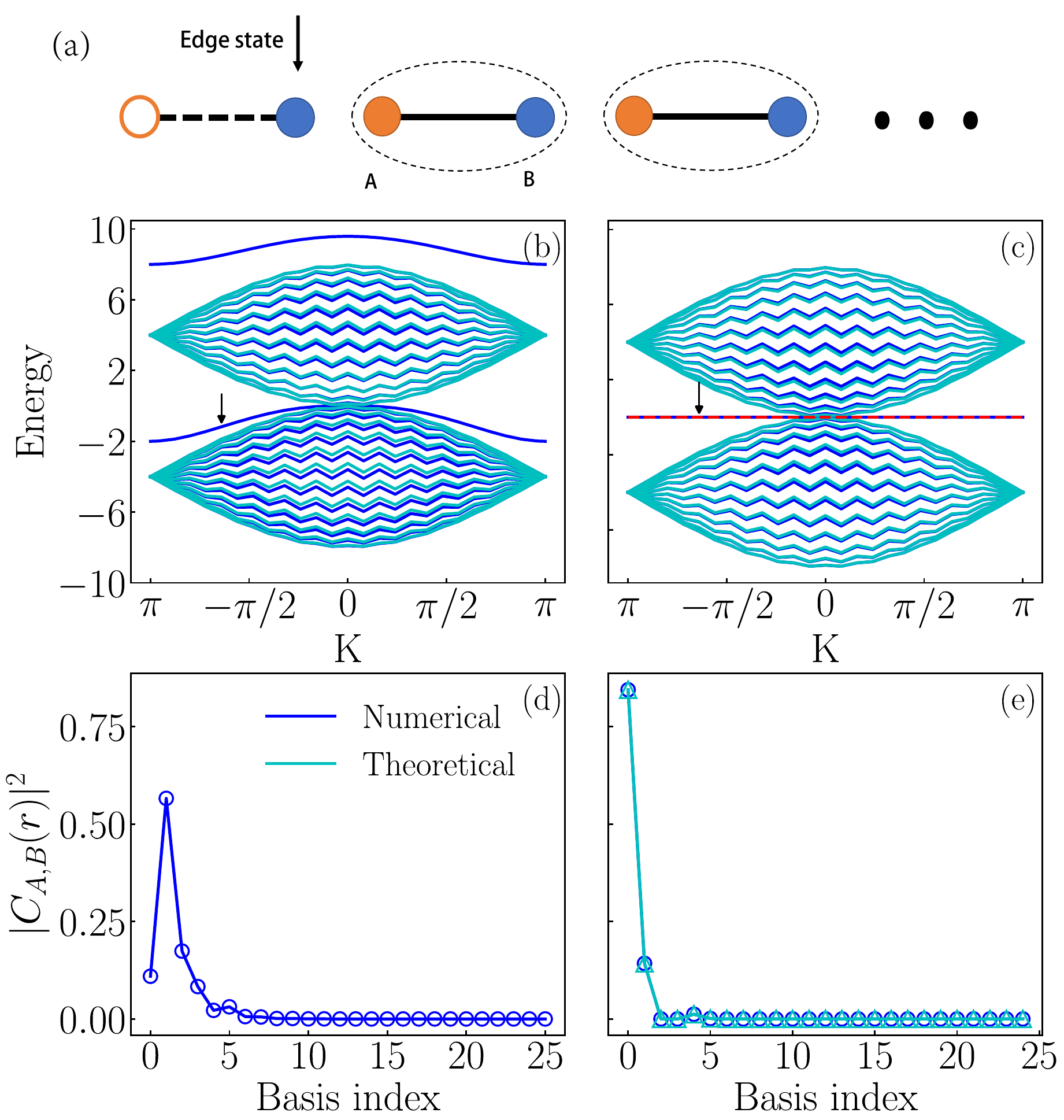}
 	\caption{(a) Sketch of the edge state for the effective Hamiltonian $\hat H_{\text{eff}}$. (b) Spectrum of two-particle system for $J_{\parallel}=1.0$, $J_{\perp}=2.0$, $U=5$, and $L=50$.
 		The major two continuous bands are determined by Eq.~(\ref{sol}).
 		The middle single mode is the edge state, and the up isolate mode is repulsively bound pair~\cite{winkler}.
 		(c) Spectrum of two-particle system for $J_{\parallel}=1.0$, $J_{\perp}=2.0$, $U=\infty$, and $L=50$.
 		The major band structure are the same as (b),
 		while there is no mode of repulsively bound pairs, and the edge state is zero-energy (red dashed line).
 		(d,e) The edge states with $K=-2.01$ ($n=17$) annotated in (b) and (c) by black arrow, respectively. The horizontal axis are index of basis arranged in $C^{K}_A(0),C^{K}_B(0), C^{K}_A(1), C^{K}_B(1),\dots, C^{K}_A(N_0), C^{K}_B(N_0)$ with $C^{K}_A(0)$ neglected in (e). The numerical results are obtained by diagonalizing Hamiltonian (\ref{Heff}) directly and the theoretical one is the solution Eqs.~(\ref{sol}) and (\ref{wf}).}
 	\label{fig:fig3}
 \end{figure}

According to Eqs.~(\ref{basis}) and (\ref{Heff}), we can find that the edge mode $|0\rangle_{B}$ of effective $\hat H_{\mathrm{eff}}$  is nothing but the rung pairs.
Hence, to reveal the rung-pair localization, it is necessary to calculate the edge state of $\hat H_{\mathrm{eff}}$.
Since the winding number is zero, generally, there is no topologically protected edge state, and the nearest-neighbor $|r\rangle_{A}$ and $|r\rangle_{B}$ can form a dimer for the ground state~\cite{su}.
Nevertheless, when the original Hamiltonian $\hat H$ is in hard-core limit leading to the absence of $|0 \rangle_{A}$, the edge dimer between $|0\rangle_{A}$ and $|0\rangle_{B}$ is broken.
Thus, a zero-energy state emerges with $|r\rangle_{\mu}$ localized at the edge $r=0$, see Fig.~\ref{fig:fig3}(a).
 	The localization in coordinate '$r$'  ensures that  $\ket{\varphi_{0,B}(K)}$ is almost the eigenstate of the Hamiltonian. However,  this is not sufficient to result in the localization of rung pair in real space. Another significant condition is that the band composed by $\ket{\varphi_{0,B}(K)}$ is flat, which means that the corresponding eigenvalue of $\ket{\varphi_{0,B}(K)}$ for different $K$ is identical. Since the rung pair state $\ket{\text{SRP}_i}$ is a superposition of $\ket{\varphi_{0,B}(K)}$ with different $K$, it is also closed to the  eigenstate of $\hat H$, i.e., the rung pair is localized in real space.
 However, there also exist propagation practically, for that $\ket{\varphi_{0,B}(K)}$ is not the strict zero-mode. Hence, the component states in the continue spectrum still propagate freely  (see Fig.~\ref{fig:fig} in Appendix.~\ref{appdB}).
In Figs.~\ref{fig:fig3}(b-e), we present the spectrum of this system with $U=5$ and $\infty$, respectively. We can find that there exists the edge modes for both cases.  However, the edge mode for $U=5$ is not flat, which makes $\ket{\text{SRP}_i}$ away from the  eigenstate of $\hat H$. This is why the localization of rung pair can only exist  in hard-core limit.
We also plot the wave functions of edge modes with $K=-2.01$ ($n=17$), and it is indeed localized at $|0\rangle_{B}$ for both cases, see Figs.~\ref{fig:fig3}(d,e).

Now we focus on the zero mode of $\hat H_{\mathrm{eff}}$.
Solving the equation $\hat H_{\mathrm{eff}}\ket{\psi(K)}=0$ approximately in hard-core limits, we can obtain the wave function as
\begin{equation}\label{wf}
\begin{aligned}
&C^{K}_{B}(0) =\frac{1}{\sqrt{2}}, \ \ \  \quad C^{K}_{B}(2 m-1)=C^{K}_{A}(2 m)=0, \\
&C^{K}_{B}(2 m) =\rho^{2 m}, \quad C^{K}_{A}(2 m-1)=\rho^{2 m-1},
\end{aligned}
\end{equation}
where $m\geq1 $ and
\begin{equation}\label{solr}
\rho={-J_{\perp}}/Q^{K}_1 + \sqrt{{(J_{\perp}}/{Q^{K}_1})^2-1}.
\end{equation}
These solution are plotted in Fig.~\ref{fig:fig3} (e), which is consistent with numerical results. Here, iff $|\rho|<1$, this zero-energy solution is a localized state at $|0\rangle_{B}$.
Solving this inequation, we obtain $|J_{\perp}|>|Q^K_1|$, i.e., $|\bar{J}_{\perp}|>2\cos(K/2)$, which is the condition of no crossing between two bands in Eq.~(\ref{sol}). Due to $K/2\in [0,\pi]$, this condition can be always satisfied for some of $K$ as long as ${J}_{\perp}\neq 0$,
and there are more localized modes as the increase of $|\bar{J}_{\perp}|$ ($|\bar{J}_{\perp}|<2$).
On the one hand, since the single rung pair state $\ket{\text{SRP}_i}$ is the linear superposition of $|\varphi_{0, B}(K)\rangle$ for different $K$, the localized strength will become stronger when there are more  localized $|\varphi_{0, B}(K)\rangle$ modes.
One the other hand, the localized length $\xi \propto -1/\ln |\rho|$ becomes smaller when enlarging $|\bar{J}_{\perp}|$.
Therefore, when increasing the reduced inter-leg hopping strength $|\bar{J}_{\perp}|$, the localization will become stronger.
For the edge effect of this localization, in Ref.~\cite{ye}, we have provided a  phenomenological description, which can interpret the boundary effect roughly.

\begin{figure*}[t]
	\centering
	\includegraphics[width=1.0\textwidth]{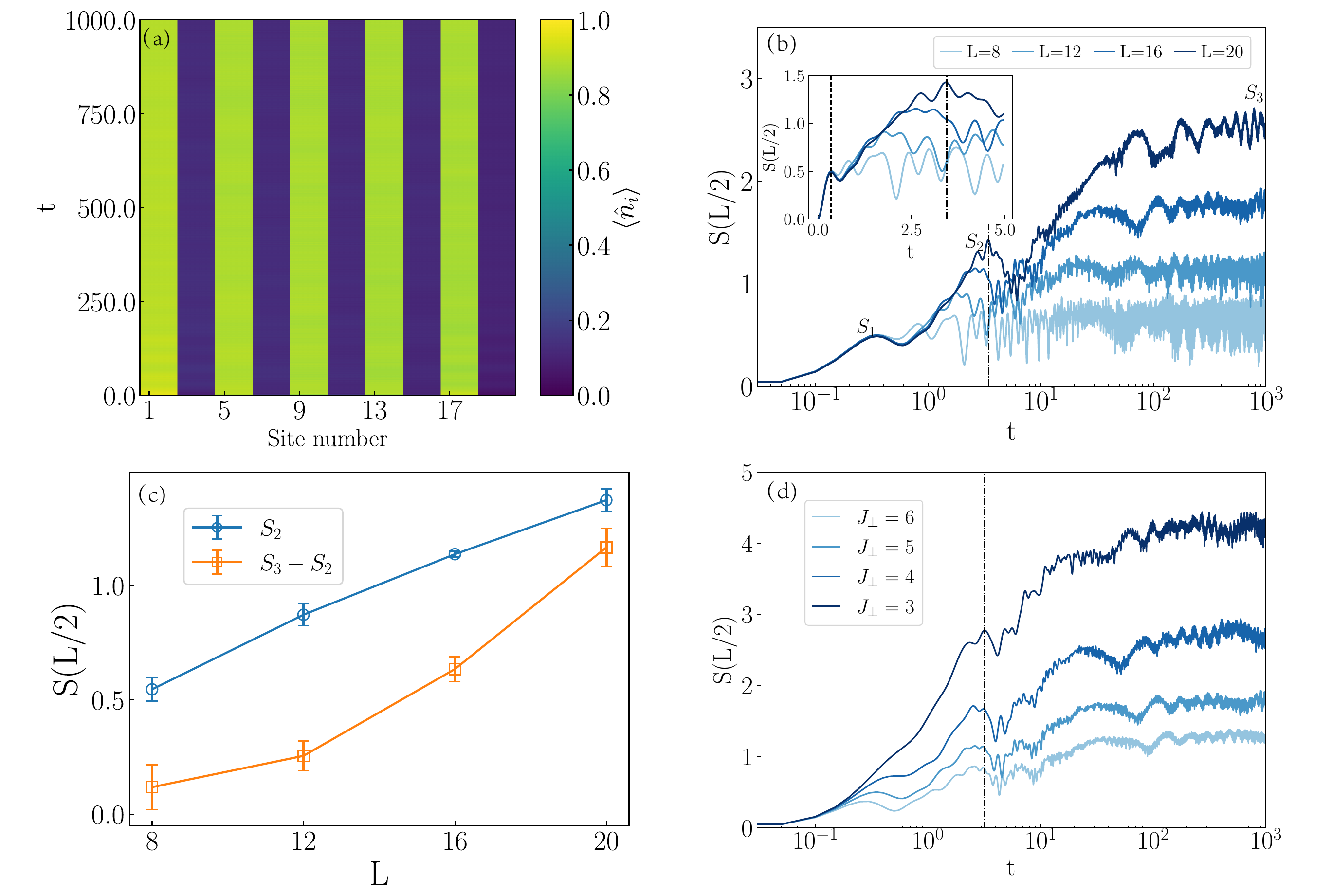}
	\caption{Dynamics of half-filling hard-core bosons. (a) Time evolution of the density distribution $\langle\hat n_{i}(t)\rangle$  with $J_{\parallel}=1.0$, $J_{\perp}=5.0$ and $L=20$. (b) Entanglement entropy growth with $J_{\parallel}=1.0, J_{\perp}=5.0$ versus different system size. The vertical dash and dot dash lines divide the dynamics into three stages (here we only mark out for $L=20$ specially). (c) Volume law of entanglement entropy in second and third stage. $S_2$, $S_3$ are the saturated values at the end of second, third stage respectively. (d) Entanglement entropy growth versus different rung hopping strength $J_{\perp}$, where $J_{\parallel}=1.0$.
	}
	\label{fig:fig4}
\end{figure*}

We note that the localization mechanism is distinct from quasi-MBL~\cite{yao}, since we cannot map the system to be a mixture of heavy and light particles.
In contrast, the rung-pair localization here is similar to the flat-band localization. However, there are two main differences between our system and the conventional flat-band model. The flat band in our system, i.e., the zero mode, is induced by strong on-site interaction and can exist as long as $J_\perp\neq0$, while the conventional flat band generally exists in the fine-tuning non-interactional system. ~\cite{mukherjee,kuno,torma,takayoshi, mondaini,yang, kobayashi}.

\section{Multiple rung pairs}\label{sec5}
 We have studied the localization of single rung pair in the viewpoint of single-particle picture.
One can verify that there exists the interaction between the rung pairs in Hamiltonian~(\ref{eq:H}) in the hard-core limit, so that the single-particle picture may be  invalid for the system with  multiple particles.
It is interesting to pursue whether the localization can still persist in the case of multiple rung pairs.
Here, we study the long-time dynamics of multi-pairs system numerically under the hard core limit and open boundary condition. When $L\leq16$, the results are obtained by the exact diagonaliztion (ED) method, while the time-evolving block decimation (TEBD) method~\cite{vidal} with second-order Trotter-Suzuki decomposition is applied for $L=20$. We set time step $\delta=0.05$ and maximum bond dimension $\chi=1200$, where the truncation error can reach $10^{-10}$. More details for TEBD are given in  Appendix.~\ref{appdB}. We have check that these error control parameters are sufficient for the convergence of time evolution.

We firstly consider the case of half filling, where each odd rung is initially occupied by a rung pair, i.e., the initial state reads $|\psi_0\rangle=\prod_{j=\text{odd}}\hat a_{j, l_1}^{\dagger}\hat a_{j,l_2}^{\dagger}\ket{\text{Vac}}$.
In Fig.~\ref{fig:fig4}(a),  we present the time evolution of the density distribution $\langle\hat n_{i}(t)\rangle$ with $\bar J_{\perp}=5$ and $L=20$,  and one can find that the localization can still exist.
Moreover, to further study the dynamics of multiple rung pairs, we calculate the entanglement entropy
\begin{equation}
S(L/2) =-\text{Tr}(\hat\rho_{L/2} \log_2   \hat\rho_{L/2}),
\end{equation}
where $\hat\rho_{L/2}$ is the density matrix of left half system, see Appendix~\ref{appdB}.
As shown in Fig.~\ref{fig:fig4} (b), one can find that the spreading of entanglement entropy consists of three stages. The first stage is a microscopic timescale relaxation dynamics of localization pair producing area-law entanglement~\cite{bardarson, huse}.
 In the second stage,
the entanglement entropy displays a linear growth, and the saturated value, labeled as $S_2$, satisfies a volume law,  see Fig.~\ref{fig:fig4}(c).
Furthermore, according to Fig.~\ref{fig:fig4}(d), we can find the spreading speed of entanglement entropy at this stage is almost $J_\perp$-independent. These behaviors are consistent with the existence of a propagation part for each rung pair, which has been discussed in Sec.~\ref{sec4}. Therefore, one can conjecture that it is the propagation part that dominates the entanglement growth in the second stage.
In addition, similar to single particle case, as  the increase of  $\bar{J}_{\perp}$, the localization becomes stronger and $S_2$ becomes smaller.

\begin{figure}[t]
	\centering
	\includegraphics[width=0.48\textwidth]{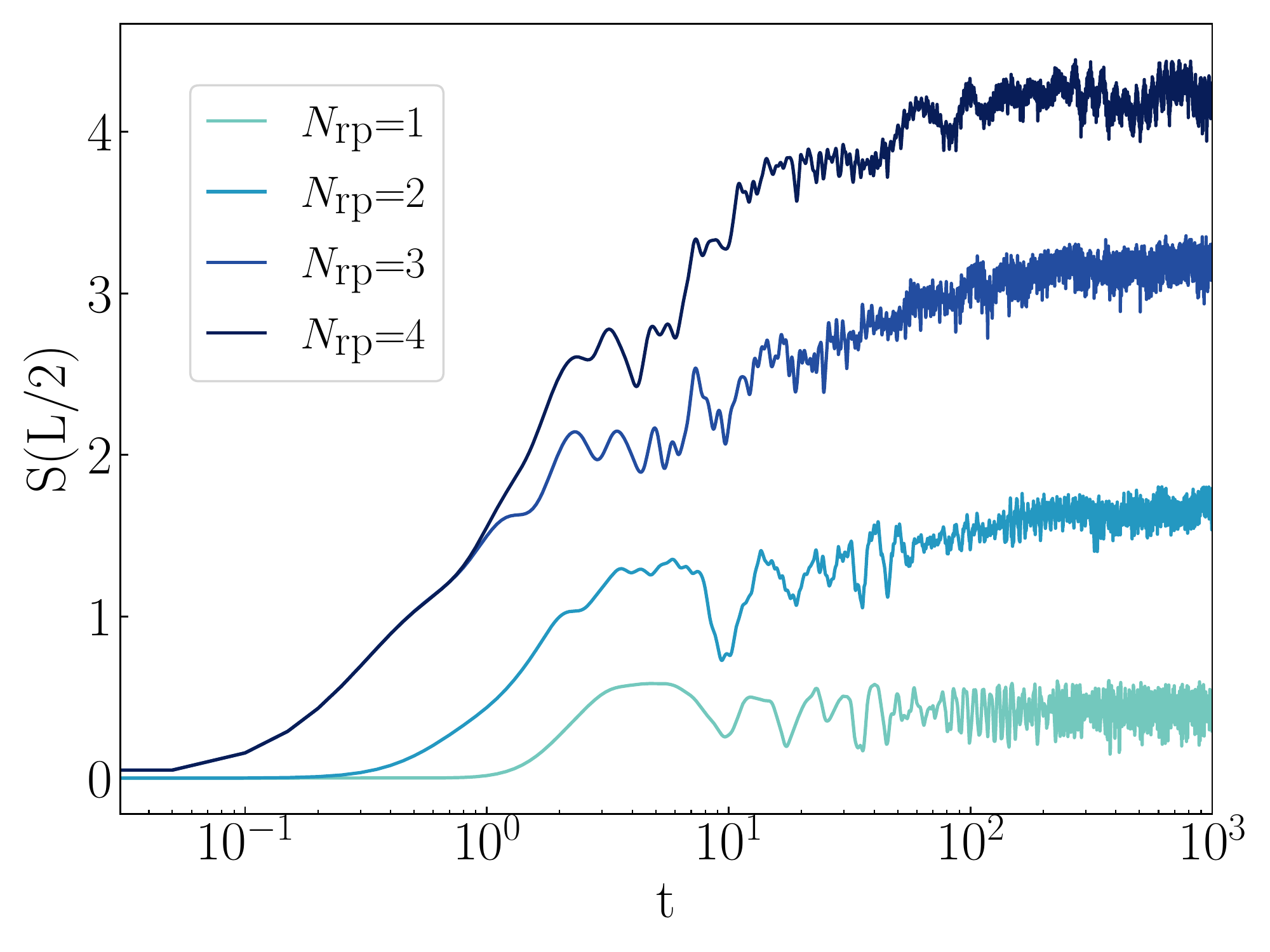}
	\caption{Entanglement entropy growth versus different number of rung pairs $N_{\text{rp}}$, where $L=16$, $J_{\parallel}=1.0$ and $J_{\perp}=3.0$. The initial rung pairs are on the first $N_{\text{rp}}$ odd rungs.
	}
	\label{fig:fig5}
\end{figure}

After the end of second stage, exotically, the spreading of entanglement entropy does not stop. Instead,  it continues to grow logarithmically for long time and finally reaches a saturated value labeled  $S_3$.
Furthermore, by subtracting $S_2$ from final saturated entropy $S_3$, we can find that the entanglement growth at the last stage, i.e., $S_3-S_2$, also tends to a volume law, see Fig.~\ref{fig:fig4} (c).
Generally, for the single-particle localized system, like AL, since each localized mode is decoupled to each other during the quench dynamics, the entanglement spreading should be only contributed by the localized modes at the boundaries of the corresponding sub-system.
That is, the growth of entanglement entropy should satisfy an area law.
Thus, in this system, the volume law of $S_3-S_2$ shows that the interaction among multiple rung pairs should contribute to the dynamics.
To further illustrate this point, we then consider the entropy spreading with different rung-pair number $N_{\text{rp}}$  for $L=16$.
In Fig.~\ref{fig:fig5}, it is shown that in the case of one rung pair, there is no logarithmic entanglement growth.
This is another evidence that the interaction can indeed affect the dynamics of multiple rung pairs, although it cannot destroy the localization.

\section{conclusion}\label{sec6}
In summary, we have studied the localizations of both single and multiple rung pairs in Bose-Hubbard ladder model.
This localization can only exist in the case of hard-core limit and becomes stronger with the increase of inter-leg hopping strength.
We map the two-particle system into an effective two-band model with approximate sub-lattice symmetry, which origins from the specific ladder structure. The rung-pair localization is related to the zero-energy mode of this effective Hamiltonian resulting from the hard-core limit.
Moreover, we also study the dynamics of multiple rung pairs and show that the localization still exists. In addition, we find that there is a long-time logarithmic growth of entanglement entropy, which is a many-body effect.
Our results reveal the mechanism of rung-pair localization in Bose-Hubbard ladder, which has been observed in quantum-simulation experiment~\cite{ye}.
This is another type of interesting disorder-free flat-band localization induced  by on-site interaction and specific lattice symmetry, which is robust with respect to interaction among particles.

Finally, there also remains several interesting open questions.
For instance, the mechanism of logarithmic growth in multi-pair case is still unclear. The thermalization properties of this system is also an interesting issue to be further studied.
Since the localized states only exist in hard-core limits, where the system is identified with spin-$\frac{1}{2}$ XX ladder model~\cite{sun, dag, iadecola, znidaric}, one may find such localized state in other spin ladder systems.

\begin{acknowledgements}
This work was supported by National Key R \& D Program of China (Grant Nos. 2016YFA0302104 and 2016YFA0300600), National
Natural Science Foundation of China (Grant Nos. 11774406 and 11934018), Strategic
Priority Research Program of Chinese Academy of Sciences (Grant No. XDB28000000),
and Beijing Academy of Quantum Information Science (Grant No. Y18G07).
\end{acknowledgements}

\appendix
\section{Derivation of two-particle spectrum}\label{appdA}
From expansion of state $|\psi(K)\rangle$ which have center-of-mass momentum $K$ in the two-particle basis $\left|\varphi_{r, A}(K)\right\rangle$ and  $\left|\varphi_{r, B}(K)\right\rangle$, i.e., Eq.(\ref{eq:psi}),
the Schr\"odinger equation Eq.~(\ref{eq:scheq})  can be written as a set of equations of $C^{K}_{A}(r), C^{K}_{B}(r)$
\begin{equation}\label{eigeneq}
\begin{aligned}
C_{A}(r+1) Q^{K}_{r}+C_{A}(r-1) Q^{K}_{r-1}+ \\
\left[(-1)^{n} Q^{K}_{r} \delta_{r, N_{0}}+U \delta_{r, 0}-\varepsilon_{K}\right] C_{A}(r)+2 J_{\perp} C_{B}(r) &=0, \\
C_{B}(r+1) Q^{K}_{r}+C_{B}(r-1) Q^{K}_{r-1}+ \\
\left[(-1)^{n} Q^{K}_{r} \delta_{r, N_{0}}-\varepsilon_{K}\right] C_{B}(r)+2 J_{\perp} C_{A}(r) &=0,
\end{aligned}
\end{equation}	
where  $\varepsilon_{K}$ is  eigenenergy for certain $K$ subspace. Here, the factors $Q^{K}_{r}$s satisfy $Q^{K}_{-1}=Q^{K}_{N_0+1}=0$, $Q^{K}_{0}=2 \sqrt{2} J_{\|} \cos (K / 2)$, $Q^{K}_{0<r \leqslant N_{0}}=2 J_{\|} \cos( K / 2)$, and  $N_0\equiv \frac{N-1}{2}$.
Thus, according to Eq.~(\ref{eigeneq}), we can obtain the effective Hamiltonian, i.e., Eq.~(\ref{Heff}). Furthermore, for convenience in numerical calculation, Eq.~(\ref{Heff}) can be written in matrix form. With the basis arranged as $C^{K}_{B}(0),C^{K}_{B}(1),\dots,C^{K}_{B}(N_0)$,$C^{K}_{A}(0), C^{K}_{A}(1)$ $,\dots, C^{K}_{A}(N_0)$, we obtain the Hamiltonian matrix in block form

\begin{equation}\hat H_{\text{eff}}=\left(\begin{array}{cc}
\hat H_{B} & \hat H_{A B} \\
\hat H_{A B} &\hat H_{A}
\end{array}\right 
),
\end{equation}
where $\hat H_{A}, \hat H_{B}, \hat H_{AB}$ are $N_0+1$ dimensional matrices, and the off-diagonal block $\hat H_{AB}$ denote the coupling of two diagonal blocks $\hat H_A$ and $\hat H_B$. Concretely, we have

\begin{equation}\hat H_{A}=\left(\begin{array}{ccccc}
U & Q_{0} & 0 & \ldots & 0 \\
Q_{0} & 0 & Q_{1} & \ldots & 0 \\
0 & Q_{1} & 0 & \ddots & \vdots \\
\vdots & \vdots & \ddots & & Q_{1} \\
0 & 0 & \ldots & Q_{1} & (-1)^{n} Q_{1}
\end{array}\right) ,\end{equation}
\begin{equation}\hat H_{B}=\left(\begin{array}{ccccc}
0 & Q_{0} & 0 & \ldots & 0 \\
Q_{0} & 0 & Q_{1} & \ldots & 0 \\
0 & Q_{1} & 0 & \ddots & \vdots \\
\vdots & \vdots & \ddots & & Q_{1} \\
0 & 0 & \ldots & Q_{1} & (-1)^{n} Q_{1}
\end{array}\right),\end{equation}
\begin{equation}\hat H_{A B}=\left(\begin{array}{ccccc}
2 J_{\perp} & 0 & 0 & \ldots & 0 \\
0 & 2 J_{\perp} & 0 & & 0 \\
0 & 0 & \ddots & & \vdots \\
\vdots & & & & 0 \\
0 & \ldots & & 0 & 2 J_{\perp}
\end{array}\right) .\end{equation}

\begin{figure}[t]
	\centering
	\includegraphics[width=0.5\textwidth]{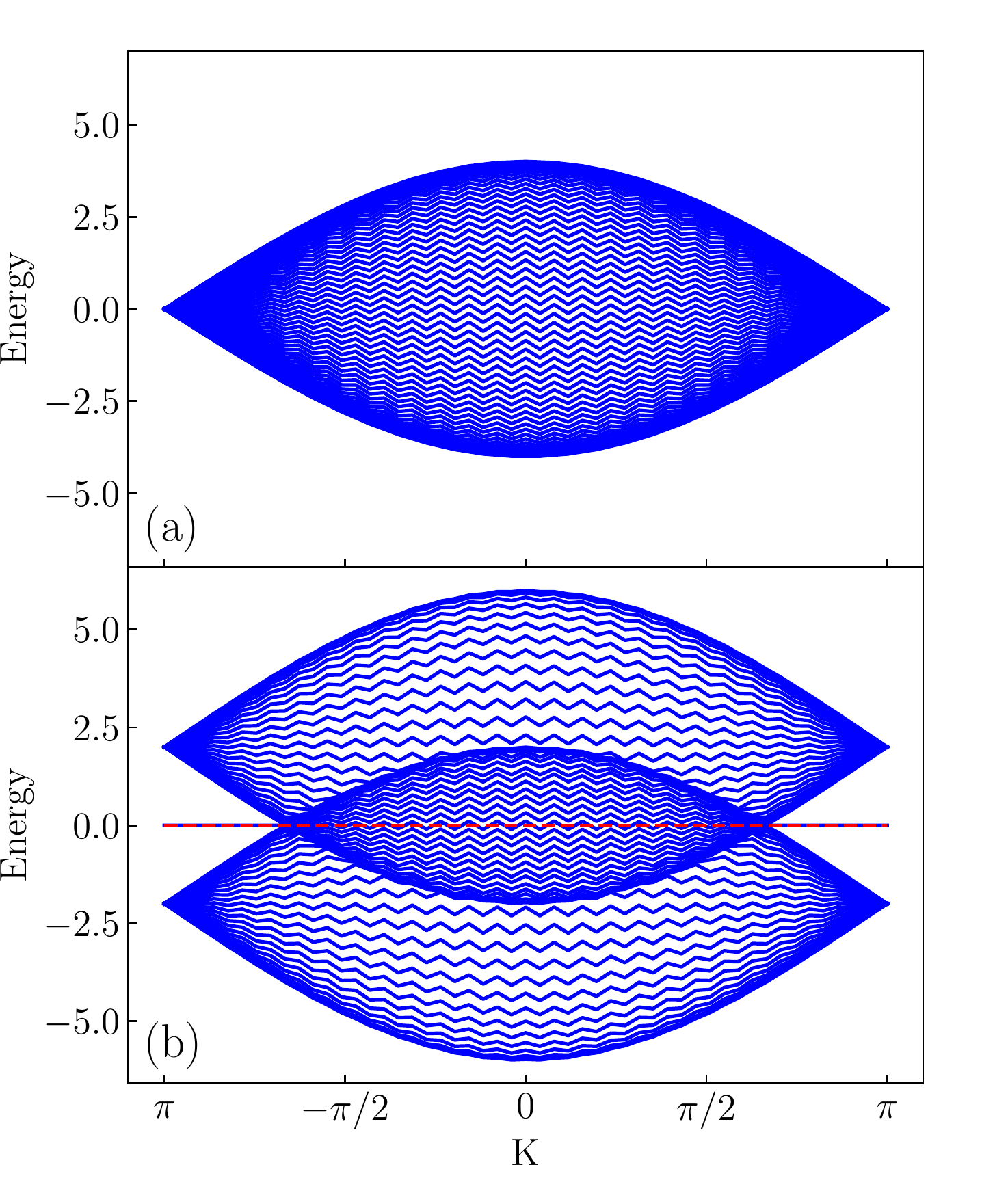}
	\caption{Two particle energy spectrum for (a) $J_{\perp}=0$ and (b) $J_{\perp}=1.0$ with $J_{\parallel}=1.0$, $U=\infty$ and $L=102$.
	}
	\label{spec}
\end{figure}

 We find that, except left boundary for $C_A(0)$, the effective Hamiltonian has a sub-lattice symmetry between $A$, $B$ sites. Diagonalizing this matrix gives the energy spectrum for the corresponding momentum $K$.  In Fig.~\ref{spec}, we plot the spectrum for $J_{\perp}=0$, $J_{\perp}=1.0$ and $U=\infty$ with site number $L=102$. It is shown that there is a band splitting and an emerging of zero mode resulted from $J_{\perp}$. We have checked that all  eigenenergies are consistent with the results obtained by exact diagonalization of $H^{\plus}$, which is the projection of Hamiltonian (\ref{eq:H}) to $+$ parity space.

The energy spectrum can also be derived analytically from solving Eq.(\ref{eigeneq}).
Because of the existence of term $U$ for $C_A(0)$, Eq.(\ref{eigeneq}) cannot have symmetry solution.
However, by combining two sets of functions, we find the combination $F^{\pm}(r)=C^{K}_A(r) \pm C^{K}_B(r)$ can have following wave ansatz form
\begin{equation}
F^{\pm}(r)=\alpha^{\pm} e^{ikr}+\beta^{\pm} e^{-ikr},
\end{equation}
and the corresponding continuous spectrum $\varepsilon_{K}^{\pm}$ are	
\begin{equation}\label{sol1}
\varepsilon_{K}^{\pm}(k)=4 J_{\|} \cos (K / 2) \cos k \pm 2 J_{\perp}.
\end{equation}
Substituting Eq.(\ref{sol1}) to the boundary equations in Eq.(\ref{eigeneq}), we obtain the constraint equations for $k$ as	
\begin{equation}\begin{aligned}
\left(2 Q_{1}^{K} \cos k-Q_{0}^{K} e^{i k}\right) \alpha+\\
\left(2 Q_{1}^{K} \cos k-Q_{0}^{K} e^{-i k}\right) \beta-U C_{A}^{K}(0) &=0, \\
\left(Q_{1}^{K}-Q_{0}^{K}\right) \alpha+\left(Q_{1}^{K}-Q_{0}^{K}\right) \beta &=0, \\
\left(Q_{1}^{K}\left[e^{i k}-(-1)^{n}\right] e^{i k N_{0}}\right) \alpha+\\
\left(Q_{1}^{K}\left[e^{-i k}-(-1)^{n}\right] e^{-i k N_{0}}\right) \beta &=0,
\end{aligned}\end{equation}	
where we find $\alpha=-\beta$ and $k$ satisfies the following relation
\begin{equation}
\sin [k(N_0+1)]-(-1)^n \sin (k N_0)=0.
\end{equation}

The wavefunction of zero mode can also be obtained in hard-core limit $U=\infty$, where the equation for $C_A(0)$ vanishes. By neglecting the boundary term and taking limit $N_0\rightarrow\infty$, we find that Eq.(\ref{eigeneq}) can be reduced to 	
\begin{equation}
\begin{aligned}
&C^{K}_{B}(2) Q^{K}_{1}+C^{K}_{B}(0) Q^{K}_{0}+2 J_{\perp} C^{K}_{A}(1) \quad=0\\
&\left[C^{K}_{B}(2 m+2)+C^{K}_{B}(2 m)\right] Q^{K}_{1}+2 J_{\perp} C^{K}_{A}(2 m+1) \quad=0\\
&\left[C^{K}_{A}(2 m+1)+C^{K}_{A}(2 m-1)\right] Q^{K}_{1}+2 J_{\perp} C^{K}_{B}(2 m) \quad=0
\end{aligned}
\end{equation}	
with $C^{K}_{B}(2 m-1)=C^{K}_{A}(2 m)=0$ and $m \geq 1$. Assuming the unnormalized solution $C^{K}_{A,B}(r)=\rho^{r},|\rho| \leq 1$, we get the solution
\begin{equation}\label{sol0}
\rho=-J_{\perp} / Q^{K}_{1} + \sqrt{J_{\perp}^{2} / {Q^{K}_{1}}^{2}-1},\quad C^{K}_B(0)=\frac{1}{\sqrt{2}}.
\end{equation}

\section{Numerical method and Numerical convergences}\label{appdB}

For numerical simulation, it is convenient to convert the ladder model to one-dimensional chain as

\begin{equation}\begin{aligned}\label{eq:1DH}
\hat{H}&=\sum_{i}^{2 N-2} J_{\parallel}\left(\hat{a}_{i}^{\dagger} \hat{a}_{i+2}+h.c.\right)+\sum_{i=\text{odd}} J_{\perp}\left(\hat{a}_{i}^{\dagger} \hat{a}_{i+1}+h.c. \right)\\ &+
\frac{U}{2} \sum_{i}^{2 N} \hat{n}_{i}\left(\hat{n}_{i}-1\right),
\end{aligned}
\end{equation}
where $N$ is the number of rung, $i$ is the site index, $\hat{a}_{2j-1}\equiv\hat{a}_{j,l_1}$, $\hat{a}_{2j}\equiv\hat{a}_{j,l_2}$.

\begin{figure}[t]
	\centering
	\includegraphics[width=0.48\textwidth=]{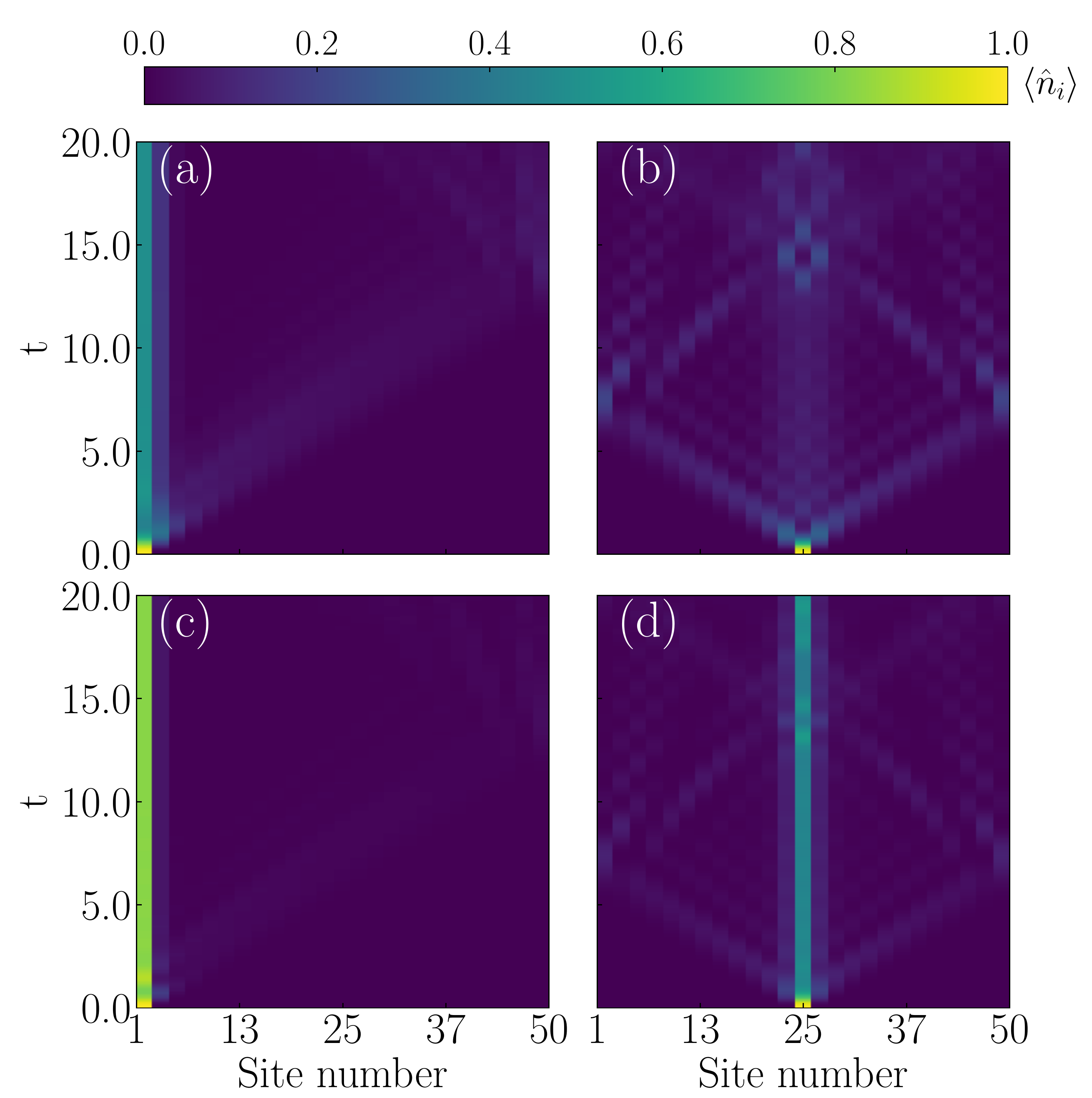}
	\caption{
		Time evolution of particle number on each site for $J_{\perp}=1.0, J_{\parallel}=1.0$ and  $U=\infty$, with the initial rung pair at (a) left edge and (b) center and that for  $J_{\perp}=2.0, J_{\parallel}=1.0$ and  $U=\infty$, with the initial rung pair at (c) left edge and (d) center.
	}
	\label{fig:fig}
\end{figure}

The two-particle dynamics can be studied by numerical simulation using ED. Figs.~\ref{fig:fig} (a), (b) show the time evolution of particle number on each site $\kb{\hat{n}_{i}(t)}$ with model parameters the same as experiment work~\cite{ye}, i.e. $J_{\perp}=J_{\parallel}=1.0$ and $U=\infty$. The site number and evolution time are $L=50$ and $t=20$, respectively. The results show that the dynamics of rung pair split to two parts: localization and propagation, which correspond to the zero-mode and continue spectrum derived in Sec.~\ref{sec4}, respectively. Also we note that the particles are still weakly localized in the bulk, which is different from the conclusion of Ref.~\cite{ye}. Similar case for $J_\perp=2.0$ are presented in Figs.~\ref{fig:fig} (c), (d), where both edge and center localization are strong enough to be observed.

\begin{figure*}[t]
	\centering
	\includegraphics[width=1.0\textwidth]{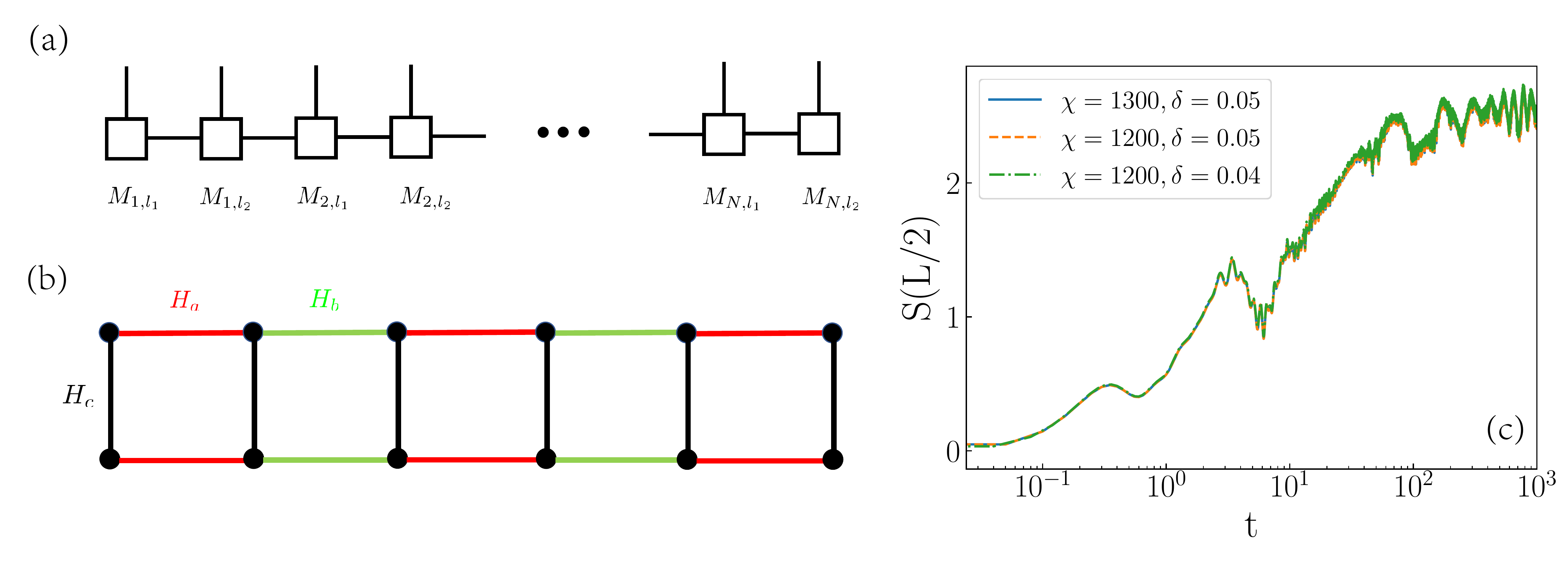}
	\caption{(a) MPS structure in TEBD simulation and bipartition of the system for calculating entanglement entropy. The entanglement entropy is obtained at central bond. (b) Decomposition of hopping terms in Hamiltonian Eq.~(\ref{eq:1DH}) for each Trotter-Suzuki step. $H_a$,$H_b$ and $H_c$ only contain local terms labeled by color red, green and black respectively. (c) Convergence of TEBD results, with the same model parameters as Fig.~\ref{fig:fig4} (b) and $L=20$. Additional parameters $\chi=1300 $ and $\delta=0.04$ are considered compared to the case of  $\chi=1200,\delta=0.05$ in the main text.
	}
	\label{figB}
\end{figure*}

 The dynamics of 20-site half-filling system are derived by TEBD method. We firstly give a brief introduction to this algorithm and then discuss the convergence of the simulation results. In TEBD,  the state is represented as matrix product states~(MPS).
\begin{equation}
\ket{\psi_{\text{MPS}}}=\sum_{s_i}M_1^{s_1}...M_L^{s_L}\ket{s_i}
\end{equation}
Here \{$s_i$\} denote the physical index for boson and $\ket{s_i}$ are the local bases. Each $M_{i}^{s_i}$ are matrix with virtual bond index except for that on the boundary, where $M_1^{s_1}$ and $M_L^{s_L}$ are in fact vectors. The dimensions of these matrices are no larger than  the maximum bond dimension $\chi$. We show the structure of MPS in Fig.~\ref{figB} (a) corresponding to Hamiltonian Eq.~(\ref{eq:1DH}). The entanglement entropy is calculated at the central bond that divide the system into half.

The time evolution of MPS state is given by second order Trotter-Suzuki decomposition of the evolution operator $e^{-i\hat{H} \delta}$ at each time step $\delta$ . The first step is to divide the Hamiltonian Eq.~(\ref{eq:1DH}) into parts that do not commute with each other. Each part contains only local terms that internally commute. For hopping term in Eq.~(\ref{eq:1DH}), the decomposition can be given by three parts, i.e., $\hat{H}_a, \hat{H}_b, \hat{H}_c$ denoted in Fig.~\ref{figB} (b). That is, $\hat{H}_a, \hat{H}_b$, and $\hat{H}_c$ only contains hopping terms labeled by color red, green and black respectively. For example,

\begin{equation}
\hat{H}_c=\sum_{i=\text{odd}}J_{\perp}\left(\hat{a}_{i}^{\dagger} \hat{a}_{i+1}+h.c. \right).
\end{equation}
 Then in each time step $\delta$, the second order Trotter-Suzuki decomposition is given by
\begin{equation}
e^{-i\hat{H}\delta}=e^{-i\hat{H}_a\delta/2}e^{-i\hat{H}_b\delta/2}e^{-i\hat{H}_c\delta}e^{-i\hat{H}_b\delta/2}e^{-i\hat{H}_a\delta/2}
\end{equation}

In the one-dimensional representation of the ladder model, there exist next-near-neighbor hopping terms, which are contained in $H_a$ and $H_b$. Thus we need to use the so called SWAP gate to exchange MPS tensor during each time step, which is discussed in Ref.~\cite{stoudenmire}. After the action of time evolution operator at each time step, the MPS must be truncated to a certain maximum bond dimension $\chi$.

The main errors in TEBD come from Trotter-Suzuki decomposition and truncations of MPS, which depend on the time step $\delta$ during the evolution and  maximum bond dimension $\chi$ of MPS, respectively. Due to the long evolution time, to ensure the results converge, we show the evolution of entanglement entropy with different bond dimensions and time steps in Fig.~\ref{figB} (c). The parameters are taken to be the same as 20-site results in  Fig.~\ref{fig:fig4} (b) , i.e. $J_{\parallel}=1.0$, $J_{\perp}=5.0$ and $U=\infty$. We find that increasing the bond dimension or decreasing the time step does not give different results. Hence our TEBD simulation is converged.

\bibliographystyle{apsrev4-1}

\begin{thebibliography}{64}%
	\makeatletter
	\providecommand \@ifxundefined [1]{%
		\@ifx{#1\undefined}
	}%
	\providecommand \@ifnum [1]{%
		\ifnum #1\expandafter \@firstoftwo
		\else \expandafter \@secondoftwo
		\fi
	}%
	\providecommand \@ifx [1]{%
		\ifx #1\expandafter \@firstoftwo
		\else \expandafter \@secondoftwo
		\fi
	}%
	\providecommand \natexlab [1]{#1}%
	\providecommand \enquote  [1]{``#1''}%
	\providecommand \bibnamefont  [1]{#1}%
	\providecommand \bibfnamefont [1]{#1}%
	\providecommand \citenamefont [1]{#1}%
	\providecommand \href@noop [0]{\@secondoftwo}%
	\providecommand \href [0]{\begingroup \@sanitize@url \@href}%
	\providecommand \@href[1]{\@@startlink{#1}\@@href}%
	\providecommand \@@href[1]{\endgroup#1\@@endlink}%
	\providecommand \@sanitize@url [0]{\catcode `\\12\catcode `\$12\catcode
		`\&12\catcode `\#12\catcode `\^12\catcode `\_12\catcode `\%12\relax}%
	\providecommand \@@startlink[1]{}%
	\providecommand \@@endlink[0]{}%
	\providecommand \url  [0]{\begingroup\@sanitize@url \@url }%
	\providecommand \@url [1]{\endgroup\@href {#1}{\urlprefix }}%
	\providecommand \urlprefix  [0]{URL }%
	\providecommand \Eprint [0]{\href }%
	\providecommand \doibase [0]{http://dx.doi.org/}%
	\providecommand \selectlanguage [0]{\@gobble}%
	\providecommand \bibinfo  [0]{\@secondoftwo}%
	\providecommand \bibfield  [0]{\@secondoftwo}%
	\providecommand \translation [1]{[#1]}%
	\providecommand \BibitemOpen [0]{}%
	\providecommand \bibitemStop [0]{}%
	\providecommand \bibitemNoStop [0]{.\EOS\space}%
	\providecommand \EOS [0]{\spacefactor3000\relax}%
	\providecommand \BibitemShut  [1]{\csname bibitem#1\endcsname}%
	\let\auto@bib@innerbib\@empty
	
	
	\bibitem{beenakker}C. W. J. Beenakker,
	Random-matrix theory of quantum transport,
	\href{https://link.aps.org/doi/10.1103/RevModPhys.69.731}{Rev. Mod. Phys. \textbf{69}, 731 (1997)}.
	
 	\bibitem{polkov}A. Polkovnikov, K. Sengupta, A. Silva, and M. Vengalattore,
	Colloquium: Nonequilibrium dynamics of closed interacting quantum systems,
	\href{https://link.aps.org/doi/10.1103/RevModPhys.83.863}{Rev. Mod. Phys. \textbf{83}, 863 (2011)}
	
	\bibitem{hasan}M. Z. Hasan and C. L. Kane,
	Colloquium: Topological insulators,
	\href{https://link.aps.org/doi/10.1103/RevModPhys.82.3045}{Rev. Mod. Phys. \textbf{82}, 3045 (2010)}.
	
	\bibitem{qi}X.-L. Qi and S.-C. Zhang,
	Topological insulators and superconductors,
	\href{https://link.aps.org/doi/10.1103/RevModPhys.83.1057}{Rev. Mod. Phys. \textbf{83}, 1057 (2011)}.
	
	\bibitem{anderson}P. W. Anderson,
	Absence of Diffusion in Certain Random Lattices,
	\href{https://link.aps.org/doi/10.1103/PhysRev.109.1492}{Phys. Rev. \textbf{109}, 1492 (1958)}.
	
	
	
	\bibitem{nandkishore}R. Nandkishore and D. A. Huse,
	Many-Body Localization and Thermalization in Quantum Statistical Mechanics,
	\href{https://www.annualreviews.org/doi/10.1146/annurev-conmatphys-031214-014726}{Annu. Rev. Condens.
		Matter Phys. \textbf{6}, 15 (2015)}.
	
	\bibitem{pal}A. Pal and D. A. Huse,
	Many-body localization phase transition
	\href{http://dx.doi.org/10.1103/PhysRevB.82.174411}{Phys. Rev. B \textbf{82}, 174411 (2010)}.
	
	\bibitem{bardarson}J. H. Bardarson, F. Pollmann, and J. E. Moore,
	Unbounded growth of entanglement in models of many-body localization,
	\href{https://link.aps.org/doi/10.1103/PhysRevLett.109.017202}{Phys. Rev. Lett. \textbf{109}, 017202 (2012)}.
	
	\bibitem{huse}D. A. Huse, R. Nandkishore, and V. Oganesyan,
	Phenomenology of fully many-body-localized systems,
	\href{https://link.aps.org/doi/10.1103/PhysRevB.90.174202}{Phys. Rev. B \textbf{90}, 174202 (2014)}.
	
	\bibitem{zangara}P. R. Zangara, A. D. Dente, P. R. Levstein, and H. M.
	Pastawski,
	Loschmidt echo as a robust decoherence quantifier for many-body systems,
	\href{https://link.aps.org/doi/10.1103/PhysRevA.86.012322}{Phys. Rev. A \textbf{86}, 012322 (2012)}.
	
	\bibitem{chandran} A. Chandran, I. H. Kim, G. Vidal, and D. A. Abanin,
	Constructing local integrals of motion in the many-body localized phase,
	\href{https://link.aps.org/doi/10.1103/PhysRevB.91.085425}{Phys. Rev. B \textbf{91}, 085425 (2015)}.
	
	\bibitem{vosk}R. Vosk and E. Altman,
	Many-Body Localization in One Dimension as a Dynamical Renormalization Group Fixed Point
	\href{https://link.aps.org/doi/10.1103/PhysRevLett.110.067204}{Phys. Rev. Lett. \textbf{110}, 067204}
	
	
	
	\bibitem{deutsch}J. M. Deutsch,
	Quantum statistical mechanics in a closed system,
	\href{https://link.aps.org/doi/10.1103/PhysRevA.43.2046}{Phys. Rev. A \textbf{43}, 2046 (1991)}.
	
	\bibitem{srednicki}M. Srednicki,
	Chaos and quantum thermalization,
	\href{https://link.aps.org/doi/10.1103/PhysRevE.50.888}{Phys. Rev. E \textbf{50}, 888 (1994)}.
	
	\bibitem{rigol}M. Rigol, V. Dunjko, and M. Olshanii,
	Thermalization and its mechanism for generic isolated quantum systems,
	\href{https://doi.org/10.1038/nature06838}{Nature \textbf{452}, 854 (2008)}.
	
	


	
	
	\bibitem{li}W. Li, A. Dhar, X. Deng, K. Kasamatsu, L. Barbiero,
	and L. Santos,
	Disorderless Quasi-localization of Polar Gases in One-Dimensional Lattices,
	\href{https://link.aps.org/doi/10.1103/PhysRevLett.124.010404}{Phys. Rev. Lett. \textbf{124}, 010404 (2020)}.
	
	\bibitem{yao}N. Yao, C. Laumann, J. Cirac, M. Lukin, and J. Moore,
	Quasi-Many-Body Localization in Translation-Invariant Systems,
	\href{https://link.aps.org/doi/10.1103/PhysRevLett.117.240601}
	{Phys. Rev. Lett. \textbf{117}, 240601 (2016)}.

	
	
	\bibitem{schiulaz}M. Schiulaz and M. Mller,
	Ideal quantum glass transitions: Many-body localization without quenched disorder
	\href{https://aip.scitation.org/doi/abs/10.1063/1.4893505}{AIP Conf. Proc. \textbf{1610}, 11
		(2014)}
	
	\bibitem{grover}T. Grover and M. P. A. Fisher,
	Quantum disentangled liquids,
	\href{https://doi.org/10.1088/1742-5468/2014/10/p10010}{J. Stat. Mech. (2014) P10010.}
	
	\bibitem{hickey}J. M. Hickey, S. Genway, and J. P. Garrahan,
	Signatures of many-body localisation in a system without disorder and the relation to a glass transition,
	\href{https://doi.org/10.1088/1742-5468/2016/05/054047}{J. Stat. Mech. (2016) 054047}.
	
	\bibitem{schiulaz1}M. Schiulaz, A. Silva, and M. Mller,
	Dynamics in many-body localized quantum systems without disorder,
	\href{https://link.aps.org/doi/10.1103/PhysRevB.91.184202}{Phys. Rev. B \textbf{91}, 184202 (2015)}.
	
	\bibitem{barbiero}L. Barbiero, C. Menotti, A. Recati, and L. Santos,
	Out-of-equilibrium states and quasi-many-body localization in polar lattice gases,
	\href{https://link.aps.org/doi/10.1103/PhysRevB.92.180406}{Phys. Rev. B \textbf{92}, 180406 (2015)}.
	
	
	\bibitem{kitaev}A. Y. Kitaev,
	Unpaired Majorana fermions in quantum wires,
	\href{https://doi.org/10.1070/1063-7869/44/10s/s29}{Physics-Uspekhi \textbf{44}, 131 (2001)}.
	
	\bibitem{su}W. P. Su, J. R. Schrieffer, and A. J. Heeger,
	Solitons in Polyacetylene,
	\href{https://link.aps.org/doi/10.1103/PhysRevLett.42.1698}{Phys. Rev. Lett. \textbf{42}, 1698 (1979)}.
	

	



     \bibitem{Smith2017_1}A. Smith, J. Knolle, D. L. Kovrizhin, and R. Moessner,
     Disorder-free localization,
     \href{https://link.aps.org/doi/10.1103/hysRevLett.118.266601}{Phys. Rev. Lett. \textbf{118}, 266601 (2017)}.	
     	
	 \bibitem{Smith2017_2}A. Smith, J. Knolle, R. Moessner, and D. L. Kovrizhin,
      Absence of ergodicity without quenched disorder: from quantum disentangled
      liquids to many-body localization,
      \href{https://link.aps.org/doi/10.1103/PhysRevLett.119.176601}{Phys. Rev. Lett. \textbf{119}, 176601 (2017)}.	
	
 	 \bibitem{Chen2018}C. Chen, F. Burnell, and A. Chandran,
     How does a locally constrained quantum system localize,
    \href{https://link.aps.org/doi/10.1103/PhysRevLett.121.085701}{Phys. Rev. Lett. \textbf{121}, 085701 (2018)}.

	\bibitem{brenes}M. Brenes, M. Dalmonte, M. Heyl, and A. Scardicchio,
    Many-Body Localization Dynamics from Gauge Invariance,
     \href{https://link.aps.org/doi/10.1103/PhysRevLett.120.030601}{Phys. Rev. Lett. \textbf{120}, 030601 (2018)}.

		
	 \bibitem{mukherjee}S. Mukherjee, A. Spracklen, D. Choudhury, N. Goldman,
	 P. \"Ohberg, E. Andersson, and R. R. Thomson,
	 Observation of a Localized Flat-Band State in a Photonic Lieb Lattice,
	 \href{https://link.aps.org/doi/10.1103/PhysRevLett.114.245504}{Phys. Rev. Lett. \textbf{114}, 245504 (2015)}.
	
	 \bibitem{kuno}Y. Kuno, T. Orito, and I. Ichinose,
	 Flat-band many-body localization and ergodicity breaking in the Creutz ladder,
	 \href{https://doi.org/10.1088\%2F1367-2630\%2Fab6352}{New J. Phys. \textbf{22}, 013032 (2020).}
	
	 \bibitem{torma}P. T\"orm\"a, L. Liang, and S. Peotta,
	 Quantum metric and effective mass of a two-body bound state in a flat band,
	 \href{https://link.aps.org/doi/10.1103/PhysRevB.98.220511}{Phys. Rev. B \textbf{98}, 220511 (2018).}
	
	 \bibitem{takayoshi}S. Takayoshi, H. Katsura, N. Watanabe, and H. Aoki,
	 Phase diagram and pair Tomonaga-Luttinger liquid in a Bose-Hubbard model with flat bands,
	 \href{https://link.aps.org/doi/10.1103/PhysRevA.88.063613}{Phys.
		 Rev. A \textbf{88}, 063613 (2013).}
	
	 \bibitem{mondaini}R. Mondaini, G. G. Batrouni, and B. Gremaud, ´
	 Pairing and superconductivity in the flat band: Creutz lattice,
	 \href{https://link.aps.org/doi/10.1103/PhysRevB.98.155142}{Phys. Rev. B \textbf{98}, 155142 (2018).}
	
	 \bibitem{yang}Z.-H. Yang, Y.-P. Wang, Z.-Y. Xue, W.-L. Yang, Y. Hu, J.-H.
	 Gao, and Y. Wu,
	 Circuit quantum electrodynamics simulator of flat band physics in a Lieb lattice,
	 \href{https://link.aps.org/doi/10.1103/PhysRevA.93.062319}{ Phys. Rev. A \textbf{93}, 062319 (2016).}
	
	 \bibitem{kobayashi}K. Kobayashi, M. Okumura, S. Yamada, M. Machida, and
	 H. Aoki,
	 Superconductivity in repulsively interacting fermions on a diamond chain: Flat-band-induced pairing,
	 \href{https://link.aps.org/doi/10.1103/PhysRevB.94.214501}{ Phys. Rev. B \textbf{94}, 214501 (2016).}
	
	\bibitem{ye}Y. Ye, Z.-Y. Ge, Y. Wu, S. Wang, M. Gong, Y.-R. Zhang,
	Q. Zhu, R. Yang, S. Li, F. Liang, J. Lin, Y. Xu, C. Guo,
	L. Sun, C. Cheng, N. Ma, Z. Y. Meng, H. Deng, H. Rong,
	C.-Y. Lu, C.-Z. Peng, H. Fan, X. Zhu, and J.-W. Pan,
	Propagation and Localization of Collective Excitations on a 24-Qubit Superconducting Processor,
	\href{https://link.aps.org/doi/10.1103/PhysRevLett.123.050502}{Phys. Rev. Lett. \textbf{123}, 050502 (2019)}.

	
	
	
	
	\bibitem{yan}Z. Yan, Y.-R. Zhang, M. Gong, Y. Wu, Y. Zheng, S. Li,
	C. Wang, F. Liang, J. Lin, Y. Xu, C. Guo, L. Sun, C.-
	Z. Peng, K. Xia, H. Deng, H. Rong, J. Q. You, F. Nori,
	H. Fan, X. Zhu, and J.-W. Pan,
	Strongly correlated quantum walks with a 12-qubit superconducting processor,
	\href{https://science.sciencemag.org/content/364/6442/753}{Science \textbf{364}, 753 (2019)}.
	
	\bibitem{song}C. Song, K. Xu, H. Li, Y.-R. Zhang, X. Zhang, W. Liu,
	Q. Guo, Z. Wang, W. Ren, J. Hao, H. Feng, H. Fan,
	D. Zheng, D.-W. Wang, H. Wang, and S.-Y. Zhu,
	Generation of multicomponent atomic {Schr\"odinger} cat states of up to 20 qubits,
	\href{http://www.sciencemag.org/lookup/doi/10.1126/science.aay0600}{Science \textbf{365}, 574 (2019)}.
	
	\bibitem{xu}K. Xu, J.-J. Chen, Y. Zeng, Y.-R. Zhang, C. Song, W. Liu,
	Q. Guo, P. Zhang, D. Xu, H. Deng, K. Huang, H. Wang,
	X. Zhu, D. Zheng, and H. Fan,
	Emulating Many-Body Localization with a Superconducting Quantum Processor,
	\href{https://link.aps.org/doi/10.1103/PhysRevLett.120.050507}{Phys. Rev. Lett. \textbf{120}, 050507 (2018)}.
	
	\bibitem{tschischik}W. Tschischik, M. Haque, and R. Moessner,
	Nonequilibrium dynamics in {Bose}-{Hubbard} ladders,
	\href{https://link.aps.org/doi/10.1103/PhysRevA.86.063633}{Phys. Rev. A \textbf{86}, 063633 (2012)}.
	
	\bibitem{keles}A. Kele\ifmmode \mbox{\c{s}}\else \c{s}\fi{}¸ and M. O. Oktel,
	Mott transition in a two-leg Bose-Hubbard ladder under an artificial magnetic field,
	\href{https://link.aps.org/doi/10.1103/PhysRevA.91.013629}{Phys. Rev. A \textbf{91}, 013629 (2015)}.
	
	
	\bibitem{vidal}G. Vidal,
	Efficient Simulation of One-Dimensional Quantum Many-Body Systems,
	\href{https://link.aps.org/doi/10.1103/PhysRevLett.93.040502}{Phys. Rev. Lett. \textbf{93}, 040502 (2004)}.
	
	\bibitem{stoudenmire}E. M. Stoudenmire and S. R. White,
	Minimally entangled typical thermal state algorithms,
	\href{https://doi.org/10.1088/1367-2630/12/5/055026}{New J. Phys. \textbf{12}, 055026 (2010)}.
	
	\bibitem{peres}A. Peres,
	Stability of quantum motion in chaotic and regular systems,
	\href{https://link.aps.org/doi/10.1103/PhysRevA.30.1610}{Phys. Rev. A \textbf{30}, 1610 (1984)}.
	
		
	\bibitem{nguenang}J.-P. Nguenang and S. Flach,
	Fermionic bound states on a one-dimensional lattice,
	\href{https://link.aps.org/doi/10.1103/PhysRevA.80.015601}{Phys. Rev. A \textbf{80}, 015601 (2009)}.
	
	\bibitem{zhang}X. Z. Zhang, L. Jin, and Z. Song,
	Non-Hermitian description of the dynamics of interchain pair tunneling,
	\href{https://link.aps.org/doi/10.1103/PhysRevA.95.052122}{Phys. Rev. A \textbf{95}, 052122
		(2017)}
	
	
	\bibitem{winkler}K. Winkler, G. Thalhammer, F. Lang, R. Grimm, J. H.
	Denschlag, A. J. Daley, A. Kantian, H. P. Bchler, and
	P. Zoller,
	Repulsively bound atom pairs in an optical lattice,
	\href{https://www.nature.com/articles/nature04918}{Nature \textbf{441}, 853 (2006)}.


    \bibitem{iadecola}T. Iadecola and M. \ifmmode \check{Z}\else \v{Z}\fi{}nidari\ifmmode \check{c}\else \v{c}\fi{},
   Exact Localized and Ballistic Eigenstates in Disordered Chaotic Spin Ladders and the Fermi-Hubbard Model,
   \href{https://link.aps.org/doi/10.1103/PhysRevLett.123.036403}{Phys. Rev. Lett. \textbf{123}, 036403 (2019).}

   \bibitem{znidaric}\ifmmode \check{Z}\else \v{Z}\fi{}nidari\ifmmode \check{c}\else \v{c}\fi{},
   Coexistence of Diffusive and Ballistic Transport in a Simple Spin Ladder,
   \href{https://link.aps.org/doi/10.1103/PhysRevLett.110.070602}{Phys. Rev. Lett. \textbf{110}, 070602 (2013)}


   \bibitem{sun} Z.-H. Sun, J. Cui, and H. Fan,
   Characterizing the many-body localization transition by the dynamics of diagonal entropy,
   \href{https://link.aps.org/doi/10.1103/PhysRevResearch.2.013163}
   {Phys. Rev. Research \textbf{2}, 013163
   	(2020).}

   \bibitem{dag} C. B. Da\ifmmode \breve{g}\else \u{g}\fi{} and L.-M. Duan,
   Detection of out-of-time-order correlators and information scrambling in cold atoms: Ladder-$\mathit{XX}$ model,
   \href{https://link.aps.org/doi/10.1103/PhysRevA.99.052322}
   {Phys. Rev. A \textbf{99}, 052322 (2019)}


%

\end{thebibliography}

%

\end{document}